\documentclass
[doublecol]
{epl2}
\usepackage{graphicx}  
\usepackage{dcolumn}   
\usepackage{bm}        
\usepackage{amssymb}   
\usepackage{amsmath}
\usepackage{color}
\usepackage{xurl}
\usepackage[utf8]{inputenc}
\usepackage[T1]{fontenc}
\usepackage[T1]{fontenc}

\usepackage{amsthm}
\usepackage{amsmath}
\usepackage{graphicx}
\usepackage{xcolor}

\usepackage{subcaption}

\makeatletter
\@ifundefined{textcolor}{}
{%
 \definecolor{BLACK}{gray}{0}
 \definecolor{WHITE}{gray}{1}
 \definecolor{RED}{rgb}{1,0,0}
 \definecolor{GREEN}{rgb}{0,1,0}
 \definecolor{BLUE}{rgb}{0,0,1}
 \definecolor{CYAN}{cmyk}{1,0,0,0}
 \definecolor{MAGENTA}{cmyk}{0,1,0,0}
 \definecolor{YELLOW}{cmyk}{0,0,1,0}
 }

\makeatletter

\newcommand{\Rmnum}[1]{\expandafter\@slowromancap\romannumeral #1@}
\makeatother

\newcommand{\be}{\begin{equation}}
\newcommand{\ee}{\end{equation}}

\def\lsim{\mathrel{\rlap{\lower4pt\hbox{$\sim$}}
    \raise1pt\hbox{$<$}}}                
\def\gsim{\mathrel{\rlap{\lower4pt\hbox{$\sim$}}
    \raise1pt\hbox{$>$}}}

\renewcommand\[{\begin{equation}}
\renewcommand\]{\end{equation}}
\usepackage[english]{babel}
\addto\captionsenglish{\renewcommand{\figurename}{FIG.}}
\makeatother

\usepackage{babel}

\begin{document}
\title{ \textcolor{black}{Active Matter as a framework for living systems-inspired Robophysics}}
\author{Giulia Janzen\inst{1},Gaia Maselli\inst{2}, Juan F. Jimenez\inst{3}, Lia Garcia-Perez\inst{3}, D A Matoz Fernandez\inst{1},Chantal Valeriani\inst{4,5}}
\institute{
  \inst{1} Department of Theoretical Physics, Complutense University of Madrid, 28040 Madrid, Spain\\
  \inst{2}  Department of Computer Science, Sapienza University of Rome, Italy\\
  \inst{3} Dept. Arquitectura de Computadores y Automatica, Facultad de Ciencias Fisicas, Universidad Complutense 28040 Madrid, Spain\\
  \inst{4}  Departamento de Estructura de la Materia, F\'isica T\'ermica y Electr\'onica, Universidad
  Complutense de Madrid, 28040 Madrid, Spain\\
\inst{5} GISC - Grupo Interdisciplinar de Sistemas Complejos 28040 Madrid, Spain\\
}

\shortauthor{Janzen \etal}

\date{\today}
\abstract{Robophysics investigates the physical principles that govern living-like robots operating in complex, real-world environments. Despite remarkable technological advances, robots continue to face fundamental efficiency limitations. At the level of individual units, locomotion remains a challenge, while at the collective level, robot swarms struggle to achieve shared purpose, coordination, communication, and cost efficiency. This perspective article examines the key challenges faced by bio-inspired robotic collectives and highlights recent research efforts that incorporate principles from active-matter physics and biology into the modeling and design of robot swarms.
}
\maketitle

\section{Introduction}
The development of robots has been a growing focus of the scientific community for over half a century, with particularly rapid advances in the past three decades \cite{Aguilar2016,Wang2021}. 
\color{black} Pioneeringly, more than two decades ago, Brooks and Flynn proposed replacing traditional expensive planetary exploration with fleets of small, autonomous, low-cost robots \cite{brooks1998fast}.
\color{black}
Such versatile devices --- remotely operated or fully autonomous --- are becoming increasingly essential, due to their ability to   perform tasks which are dangerous, repetitive, or otherwise challenging for humans.
Robots offer significant benefits in tasks such as search and rescue operations~\cite{Lyu2023}, being able to access unsafe environments, as well as damage evaluation, agricultural monitoring, forest management~\cite{ecke2022uav}, and crowd monitoring~\cite{husman2021unmanned,conte2021performance,al2017crowd,gohari2022involvement},   enhancing precision and efficiency.
Despite their remarkable capabilities and diverse applications, robots still encounter substantial efficiency limitations both as individual units and in collective operations. In comparison to biological organisms, at the single-robot level, locomotion remains a critical challenge: robots often struggle to navigate natural terrains, such as dense forests, rubble-strewn areas, or regions with gusty winds\cite{Aguilar2016,Schaeffer2013, Fajen2013, Biewener2018}. Additionally, mission planning is frequently challenging due to the robot's inability to effectively manage unforeseen obstacles, resulting in decreased operational efficiency and reliability. To address these challenges, recent research efforts have focused on incorporating principles from physics and biology into the modeling and design of robots\cite{Volpe_2025}.

The gap in understanding motion within complex environments, particularly from a physics perspective, has given rise to the field of Robophysics \cite{Aguilar2016}. Robophysics is defined as the pursuit of principles governing the movement and control of robot-like entities in intricate environments. It delves into fundamentally non-equilibrium processes where motion emerges through internal actuation \cite{Levine2023}. Advancing the understanding of robotic movement principles necessitates the application of experimental physics, combining theoretical and numerical frameworks with systematically controlled laboratory experiments on simplified systems.

\color{black}
In this perspective article, we examine how Robophysics provides a framework for understanding and improving both individual and collective robotic behaviors. We first discuss how insights from living systems have informed the design of single-robot locomotion strategies. Next, we  explore the extension of these principles to collective systems, highlighting the challenges that arise in robot swarms. Finally, we outline how recent advances in active matter physics and machine learning can be integrated to develop predictive, adaptive, and purpose-driven robotic collectives.
\section{Learning from living systems: single robot} 
At the {\it single-robot level}, Robophysics has focused on understanding and improving locomotion by building simplified devices capturing essential geometric and control features of biological systems \cite{Aguilar2016}. These devices function as abstract models, providing insights into appendage-induced locomotion, blurring the dichotomy between failure and success to uncover principles of environment-locomotor interaction \cite{Aguilar2016}. Consequently, the exploration of robotic motion serves as an exemplary model system for advancing our understanding of complex dynamical systems driven far from equilibrium \cite{bechinger2016active,Volpe_2025}.  
So far, the scientific literature has reported both terrestrial and aerial examples in which Robophysics played a pivotal role in enabling robots to adapt to real-world conditions by incorporating living-like mechanisms into their design. 

On the one side, on land, \textit{snake-like robots} have been designed to move through unpredictable and confined spaces by mimicking the undulating motion of snakes, which use their body segments and frictional forces to generate propulsion  \cite{snake1,snake2}. This ability makes snake-like robots highly suitable for deployment in disaster zones, where they can traverse rubble and debris to locate survivors or assess structural damage\cite{snake3,snake4}. Studies of snake locomotion across diverse terrains have allowed the creation of robots excelling in mobility, adaptability, and durability, even under extreme conditions \cite{MOOSAVI2025103418,volpeworm2025}. 
At a smaller length-scale, peristaltic {\it \textcolor{black}{worm-inspired}}  soft robots have been designed for a better understanding of locomotion in different environments \cite{Das2023}.

On the other side, in the air, \textit{bird-inspired drones} have been tailored to exploit flapping-wing aerodynamics to achieve maneuverability and energy efficiency which surpasses traditional fixed-wing or rotor designs\cite{Phan2024}. 
At a smaller length scale, insect-inspired robots have been designed for crop pollination and environmental monitoring\cite{Chen2019}.
Such bio-inspired abstractions demonstrate how Robophysics, by bridging biology and engineering, enhances robotic adaptability in complex real-world environments \cite{Jeger2024}. 

In what follows, we show that a similar approach can be taken at the collective level. Just as individual robots benefit from bio-inspired locomotion strategies, robot swarms can benefit from coordination and adaptability principles observed in animal groups and other living ensembles.

\section{Learning from living systems: collective behaviour} 
Living systems achieve collective functionality not via a central control, but via local interactions which generate robust and adaptable group behavior \cite{Hauser1997,Gillam2011}. Across length-scales, organisms actively consume energy to remain out of equilibrium and thereby dynamically respond to environmental cues \cite{wang2015one}.  \textcolor{black}{Ranging from the micro to the macroscale, bacteria  form dense swarms to withstand exposure to antibiotics \cite{ButlerPNAS2010}, undergoing self-sustained active turbulence relevant for fluid mixing and molecular transport \cite{DunkelPRL2013,WensinkPNAS2012,SokolovPRL2012}} 
, cytoskeletal filaments continuously assemble and disassemble to enable dynamic cell shape changes \cite{fletcher2010cell}, while flocks of birds or schools of fish reorganize within fractions of a second to conserve energy or evade predators \cite{Marchetti2013,couzin2005effective,portugal2014upwash}. The unifying principle among all these systems is that coordinated responses emerge from simple, local rules of sensing and interaction, yet giving rise to collective resilience and adaptability.

Efficient robot swarms \textcolor{black}{should} rely on fundamental processes-communication, coordination, cooperation, collaboration, and, at times, competition \cite{prorok2021}, to achieve collective goals. Algorithms based on these tasks have already achieved remarkable results in controlled experimental environments with a limited number of individuals in the swarm (fewer than ten). The following sections discuss each of these aspects and their relevance to collective robotic behavior.

\subsection{Communication and its constraints}
\color{black}
Communication is the key for enabling swarm coordination. In robophysical collectives, it is fundamentally constrained by density, mobility, and morphology, which jointly determine the timeliness, reliability, and cost of information exchange.
\color{black}
Gielis et al.~\cite{Gielis2022} argue that communication in robot swarms confronts two major challenges. First, learning models should be employed not only to determine what information to communicate and to whom, but also  optimize multiple aspects of swarm performance.  
Second, robots must be designed together with their communication capabilities, since any control algorithm relying on explicit communication is prone to failure when network data are insufficient and to sub-optimal performance when the network transmits excessive information.

\color{black}Specifically, these constraints are regime-dependent: changes in density and morphology can abruptly alter contention, connectivity, and information freshness, thereby setting hard limits on coordination performance.
For this reason, standard communication protocols cannot be adopted directly, because they aim at improving generic network performance metrics rather than ensuring the timeliness and scalability required by swarm coordination.
Consequently, robophysical swarms call for task-driven, control-aware communication that explicitly limits transmissions and adapts what/when/to whom  communicating (and at which rate) to maintain timeliness and scalability across varying density and morphology regimes.

More fundamentally, in Robophysics communication is not a passive substrate but part of the collective dynamics: information exchange affects motion and coordination, while motion continuously reshapes connectivity and contention.
Designing mechanisms that explicitly account for this bidirectional coupling---and remain stable across regime shifts---remains an open challenge.
\color{black}

\subsection{Coordination and Scalability}
\color{black} 
In robotic swarms, objectives are collective and must be achieved through strictly local interactions with limited global knowledge, making coordination a central challenge. Coordination aims to minimize redundant actions and interference through cooperative behaviors such as trajectory coordination for area coverage and collision avoidance~\cite{yao2022guiding}.
\color{black}Scalability is essential as the number of robots increases, yet many techniques  performing well for small groups become impractical at larger swarm sizes. Recent research has therefore focused on the formal control of shape formation in robot collectives~\cite{sakurama}. In these systems, robots communicate with local neighbors to maintain a desired configuration, and their interactions are typically represented as a graph. In robophysical swarms, however, this graph is not merely a design abstraction but an outcome of morphology and density; as the collective transitions across regimes, graph connectivity and the available bandwidth per node can change abruptly, challenging the stability and convergence of formation and coverage controllers. Scaling up such formations requires enlarging the interaction graph while preserving key topological properties, a task that remains particularly challenging.

\begin{figure}[h!]
\begin{center}
\includegraphics[width=0.8\columnwidth]{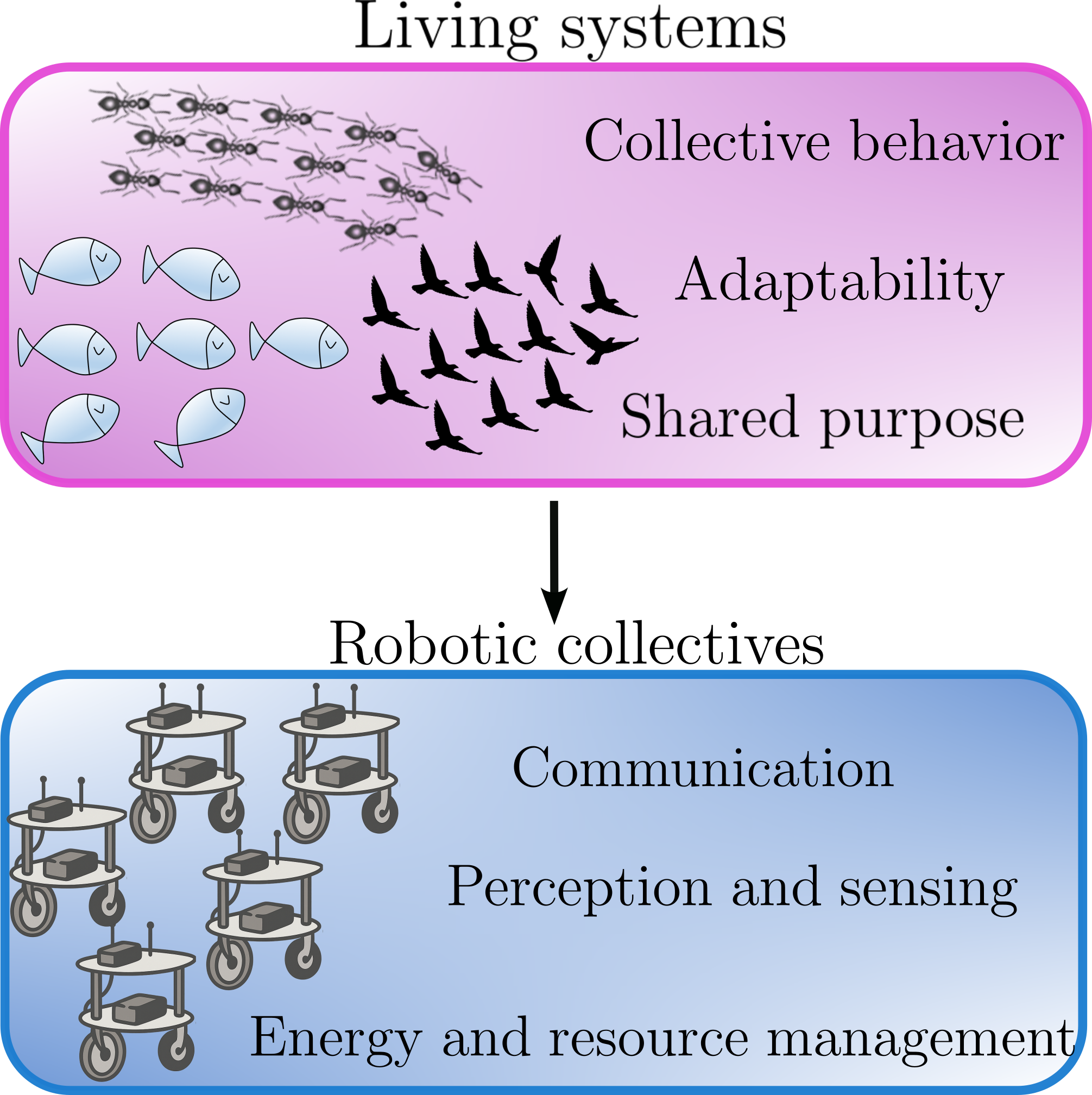}
\end{center}
\caption{Schematic illustration of how principles from living systems can inform robophysics. In living systems (top), groups of organisms such as fish, ants, and birds display collective behavior, adaptability, and shared purpose. These principles inspire robotic collectives (bottom), where multi-robot swarms must address challenges such as communication, perception and sensing, and energy and resource management.\label{fig_robots} }
\end{figure}

\subsection{Networking and cooperation}
At first glance, robophysical networking issues may appear similar to those in bio-inspired networking, as both draw on concepts from collective behavior in natural systems. However, bio-inspired networking primarily aims to leverage biological models to improve the performance of routing, coordination, synchronization and resource-allocation algorithms in large-scale communication networks~\cite{Falko}. Robophysics, instead, seeks to develop a physics-based understanding of how real robots move, interact, and generate collective behaviors under physical constraints.

From a computer networking perspective, the operation of a swarm of small, low-cost robots faces several fundamental challenges. Unlike abstract network nodes, robots are physical entities whose positions and motions directly affect network connectivity. As a result, the network topology continuously evolves with swarm motion, requiring tightly coupled strategies for locomotion, control, and communication.
\color{black}
Crucially, many swarm networking functions (e.g., neighbor discovery, local coordination, and topology-aware communication) implicitly assume  robots  estimating the identity of their neighbors and the evolution of  the interaction graph. 
Two notions are central in robophysical networking: \emph{embodiment} (robots'physical motion and morphology shape connectivity, interference, and access to shared resources) and \emph{embeddedness}(sensing, computation, and communication must be carried onboard).
The latest happens 
under strict energy and form-factor constraints and often without external infrastructure (e.g., anchors, beacons, motion-capture, or centralized fusion). Importantly, these notions are coupled: embedding additional functionality (sensors, computation, radios, batteries) changes size, mass, and power draw, thereby affecting locomotion and, ultimately, the networked collective dynamics.

In location-deprived environments (such as indoor  experiments) simultaneous localization and mapping (SLAM) becomes essential, 
Typical SLAM and localization pipelines often rely on assumptions which do not hold in robophysical swarms, such as the availability of external infrastructure (e.g., anchors), pre-installed maps, or stable reference frames. 
In fully infrastructure-free deployments, robots must instead infer relative positions and the interaction graph from sparse, noisy onboard sensing and intermittent communication.
Performing topology mapping and cooperative localization in indoor environments with only simple sensors is thus a significant challenge. Relying on low-cost sensors inherently introduces sensor uncertainty, since measurements of distance, velocity, or orientation are often noisy or incomplete. This, in turn, leads to perception uncertainty, as obstacles or neighboring robots may be only partially detected or misidentified due to occlusions or reflections. Moreover, robot mobility exacerbates communication uncertainty, as time-varying and intermittent wireless links can cause packet losses and transmission delays.

\color{black}Cooperation refers to the ability of multiple (often homogeneous) agents to accomplish tasks that would be infeasible individually, such as jointly transporting a payload or assisting one another in traversing obstacles. For decades, computer networking has drawn inspiration from collective behavior in natural systems, leading to bio-inspired paradigms such as \emph{ant-colony}-inspired routing and optimization schemes. In ant-based routing for mobile {\it ad hoc} networks, artificial ``ants'' (control or data packets) explore alternative paths and probabilistically reinforce successful ones, mimicking pheromone-mediated trail formation. This approach, known as Ant Colony Optimization (ACO), has proven effective in dynamic, decentralized settings by leveraging self-organization and adaptability to changing network topologies.

Similarly, evolutionary algorithms such as genetic algorithms (GAs) have been widely used for path planning and task assignment in multi-robot and multi-drone systems. For instance, GA-based strategies such as GenPath~\cite{GenPath} optimize UAV missions under constraints including limited battery life, intermittent communication, and multi-phase recharging, enabling scalable coordination in applications such as search-and-rescue, environmental monitoring, and delivery.

Despite their success, ACO and GA-based methods are  formulated around offline or well-defined optimization objectives and do not fully capture the real-time, embodied responsiveness of biological collectives, where coordination emerges from local physical interactions and rapid adaptation under changing conditions and constraints. This gap motivates Robophysics as a complementary viewpoint: rather than imposing top-down coordination, it seeks to program an emergent collective behavior  from the local interactions' physics, fluctuations and energy consumption.


\subsection{Collaboration, Learning and Adaptation}
Collaboration emphasizes teamwork 
leveraging the unique skills of different individuals, such as integrating various sensors or carrying out a common patrol task,  combining aerial and aquatic autonomous vehicles.

Learning concerns the swarm's ability to learn from new scenarios and adjust its collective behaviour to achieve its goals, which is crucial for effective operation.
While learning,  robots in a swarm can collectively perceive their surroundings and share that information among themselves (distributed perception and sensing). \color{black} In the paradigm of embodied evolution, for instance, robots continuously exchange and modify their own control ``genomes'' through local interactions, enabling real-time adaptation to the environment and to the collective dynamics themselves \cite{bredeche2018embodied}.
\color{black}
To better adapt to the environment, robots should exhibit both robustness and fault tolerance. 
Although these properties are related, they represent distinct concepts. Robustness refers to the ability to maintain satisfactory performance even when environmental conditions differ from those anticipated during the design phase. Fault tolerance, instead, refers to the capacity of the swarm to sustain effective operation despite failures in individual agents, sensors, or other components.
\subsection{Competition and  a shared purpose}
In both biological and robotic collectives, competition can coexist with cooperation, as individuals pursuing local goals may still contribute to global functionality. 
Such emergent coordination depends not only on feedback regulation but on a shared purpose. At the microscopic scale, gene regulatory networks couple local signals to coordinated multicellular responses during processes like morphogenesis, embodying distributed feedback \cite{grada2017research,angelini2011glass,bi2016motility}. At the macroscopic scale,  local interaction rules (e.g., alignment, attraction, repulsion) are \emph{goal-conditioned}: animals modulate the way they weigh neighbors and environmental cues in service of collective functions such as predator evasion, energy saving, or foraging \cite{couzin2005effective,portugal2014upwash,reebs2000can,swaney2001familiarity}. Robustness in living systems thus arises from \emph{distributed decision-making with a purpose}, real-time adaptability driven by local feedback tuned to a common objective, rather than pre-planned, centralized control \cite{marshall2024aims}.

Translating this principle to robotics suggests that achieving efficient, self-organizing collective motion requires embedding shared purpose directly into swarm control architectures. Incorporating goal-oriented feedback and distributed decision-making into physical design rules could enable robotic swarms to exhibit the same adaptability, resilience, and emergent coordination that characterize living systems.
\subsection{Testing robot swarms}  
To test and refine swarm technologies,  
small-scale and inexpensive testbeds are crucial\cite{mokhtarian}. 
As noted by M.~Kegeleirs \textit{et al.}\cite{Kegeleirs2025}, one of the major obstacles to real-world swarm deployment is the lack of modern, affordable, and reliable experimental platforms. A number of swarm-specific testbeds have been developed, including E-Puck \cite{mondada2009puck}, Kilobot \cite{rubenstein2012kilobot}, Duckietown \cite{paull}, and Crazyflie \cite{giernacki2017crazyflie}. These systems provide standardized, cost-effective environments for studying collective behavior at small scales.

Nevertheless, a common alternative is to build an in-house platform with custom-built \cite{garcia-perez} or commercial robots \cite{10092576}, 
\color{black}{such as those built at the Georgia Tech Robotarium \cite{wilson} or at UCM-Robotarium. Figure \ref{fig_robots} shows a group of five robots (UCM-Robotarium), each with an Arduino processor  controlling the motors' speed, receiving feedback signals from the wheel encoders. }
\begin{figure}[h!]
\begin{center}
\includegraphics[width=0.6\columnwidth]{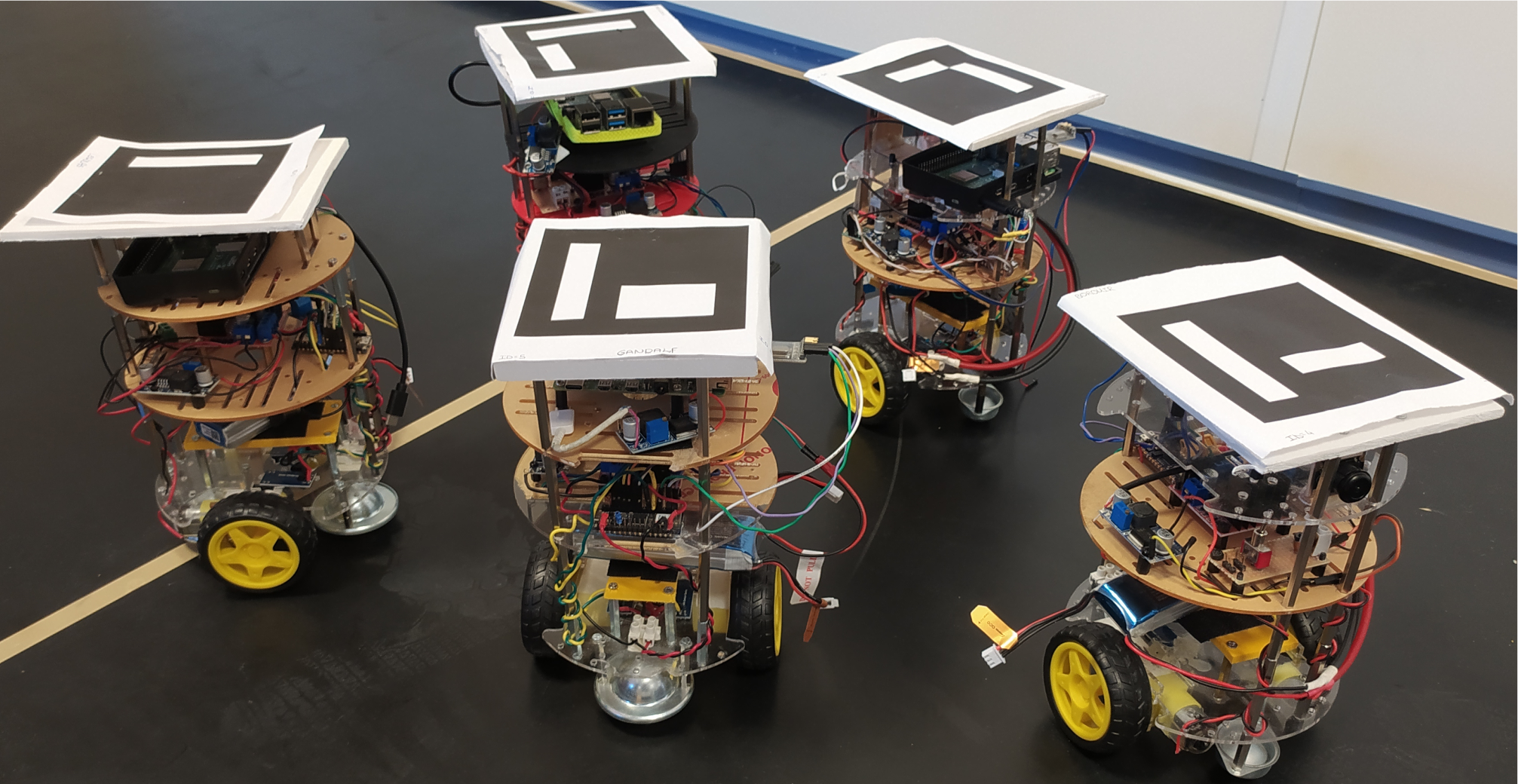}
\end{center}
\caption{Group of five differential robots from the  Robotarium-UCM.  \color{black}{The Arduino communicates with a Raspberry Pi via a serial port. This Raspberry Pi receives the position or orientation of the robot relative to the arena robot from a server, thanks to an ArUCo-type marker. Some of the robots have battery sensors, and others have infrared distance sensors.}\label{fig_robots} }
\end{figure}

\color{black}
Finally, efficient energy and resource management are critical for the long-term operation of swarm systems. Developing strategies to monitor and balance power consumption across agents is essential for scaling swarm robotics toward real-world, autonomous applications.
\color{black}
These challenges can be naturally organized according to their functional role and level of dependency. At a foundational level, the availability of experimental platforms, scalability and decentralization, the cooperative localization in GPS-deprived environments, the coupling between communication and motion, and the energy/resources management are essential prerequisites for the operation of large robotic swarms. Building upon this physical and operational substrate, higher-level challenges are the emergence of coordinated collective behaviors, including  a shared purpose and goal-oriented dynamics. Finally, adaptive and resilience-related aspects such as learning, robustness, and fault tolerance represent advanced capabilities that can be progressively developed once a coherent and scalable collective system is in place.

Notably, the foundational and coordination-level challenges are precisely those for which the Robophysics and active matter frameworks provide particularly powerful conceptual and experimental tools, as they explicitly address the coupling between physical interactions, motion, communication, and emergent collective dynamics.

\color{black}
\section{Active Matter as an effective  framework for Robophysics}
\textcolor{black}{ Intelligent Soft Matter has been recently developed as a field of research aimed at creating biological-inspired materials with life-like capabilities, such as perception, learning, memory, and adaptive behavior\cite{baulin2025intelligent}. }
Active matter has emerged in recent years as a robust framework for studying collective behaviors in both living\cite{Marchetti2013,bechinger2016active,Gompper2025} and synthetic\cite{JuACSnano2025} systems. It focuses on ensembles of self-propelled agents, whether biological (e.g., bacteria, fish) or artificial (e.g., colloids),  consuming energy to generate motion and stresses, thereby driving the system out of thermodynamic equilibrium \cite{Marchetti2013}. These agents can exhibit striking emergent phenomena, such as flocking or swarming,  purely arising  from local interactions. 
\textcolor{black}{Synthetic micro/nanorobots could have potential applications in biomedicine and 
environmental remediation.\cite{JuACSnano2025}}
\textcolor{black}{Flocking  emerge in suspensions of self-propelled Janus colloids with anisotropic repulsioin at the front/rear 
 characterised by turn-away torques
 \cite{DasPRX2024}. Light-driven synthetic micro-swimmers  form dense swarms capable of trapping and transporting passive loads  \cite{BenZion2022NAtureComm}, whereas dense binary mixtures of active Janus colloids (with distinct motilities and independently tunable alignment \cite{alvarez2025segregation}) aggregate into self-organized and highly dynamic polar clusters. }
\color{black}
In contrast to biological systems, most active matter models lack internal decision-making and goal-oriented dynamics: while  capturing emergent motion, they rarely encode  shared purposes that living collectives achieve, such as predator evasion, energy optimization, or efficient foraging \cite{te2024review}. 
Recent developments highlight ongoing efforts to close this gap. New models seek to incorporate perception, adaptability, and purposeful behavior into synthetic active systems \cite{MarchPons2025,MarchPonsEtAl2024,AcostaSobaEtAl2025,AbdoliSharmaLowen2025,lowen2025towards,cai2025reinforcement}.
\color{black}{Cooperative transport   spontaneously emerges  in swarms of stochastic self-propelled robots \cite{Arbel2025}
and in 
swarms of robots\cite{CasiulisPNAS2025}, with applications  in embodied decentralized control.
Swarms of  active particles confined in a flexible scaffold have been proposed  to build   functional soft structures. Mindless rod-like motile robots in confinement  undergo collective behavior near  boundaries\cite{Deblais2018PRL}, and are capable to perform simple tasks such as pulling a load \cite{Boudetscience2021}.}
\color{black}

At the same time, machine learning methods have been nowadays increasingly applied to active matter systems\cite{bapst2020unveiling,boattini2021averaging,clegg2021characterising,WetzelEtAl2025,JungEtAl2025,janzen2024classifying,CichosEtAl2020,janzen2023dead,volpebook2024}. In particular, neural networks have been integrated into numerical simulations and trained on experimental data, enabling  simulations to capture complex, data-driven cell behaviors \cite{minartz2025deep}. Neural network classifiers have been trained on numerical  data to distinguish motile from non-motile cells \cite{braat2025shape}, and graph neural networks have been used to extract the underlying pairwise interaction  governing the dynamics of active colloids from experimental data \cite{ruiz2024discovering}.

Beyond analysis, machine learning methods have been applied for \textit{inverse design}\cite{liu2025machine},  seeking to identify microscopic ingredients or control parameters leading to a desired macroscopic response. 
\textit{Reinforcement learning}\cite{sutton1998reinforcement,brunton2022data} (RL) is particularly  suited for this task, as it mimics trial-and-error strategies typical of biological systems. In Reinforcement Learning, an agent interacts with its environment, receiving positive or negative feedback that guides it toward actions maximizing a long-term reward \cite{brunton2022data}.
Such approaches have already been applied to soft and active matter systems to discover optimal protocols for steering active particles between states at minimal energetic cost \cite{PhysRevResearch.3.033291}, to dynamically control  self-assembly of quasicrystalline structures \cite{lieu2025dynamic} and to learn control strategies for non-equilibrium dynamics using only local
gradient information \cite{floyd2024learning}. \textcolor{black}{Recent work has also brought RL and related adaptive feedback schemes into a physical setting of mechanically interacting robotic swarms, demonstrating adaptive phototaxis via positive and negative feedback self-alignment \cite{Mirhosseini2022} and morphological computation with decentralized learning in sterically interacting robots \cite{BenZion2023}.}

\section{Outlook}
Looking forward, we believe that active matter physics can be considered as the framework to tackle Robophysics. Active Matter allows  to extract the minimal ingredients needed to reproduce a Nature-inspired collective motion including purpose, perception, communication, and adaptive feedback, and to translate all into robotic swarms.


Since machine learning and reinforcement learning have recently proven to be powerful tools for analyzing and controlling complex dynamical systems, their integration with active matter offers a path toward a conceptual framework that identifies the essential ingredients of living collectives and enables their implementation in robotic systems. Achieving this goal will require an interdisciplinary approach: physics to identify the governing principles, robotics and engineering to implement them in hardware, and computer science to provide the communication and data-processing infrastructure. By merging these perspectives, active matter will leverage robophysics to move beyond locomotion toward programmable collectives that approach the functionality of living systems.

\section{Acknowledgments}

C.V. acknowledges fundings
IHRC22/00002 and Proyecto PID2022-140407NB-C21 funded by
MCIN/AEI /10.13039/501100011033 and FEDER, UE.
DMF and GJ acknowledge support from MINECO (MCIN/AEI Grant No. PID2023-148991NA-I00). DMF also is funded through the Ramon y Cajal Program of the Spanish Ministry of Science and Innovation (Grant No. RYC2023-043002-I). J.J. and L.G-P. acknowledge support from iRoboCity2030-CM, Robotica Inteligente ($TEC-2024/TEC-62$), funded by the Programas de Actividades $I+D$ en Tecnologias en la Comunidad de Madrid. G.M acknowledges fundings Sapienza Award Horizon Europe - 2024.

\bibliographystyle{eplbib}
\bibliography{biblio}

@article{Schaeffer2013,
  title={How animals move: comparative lessons on animal locomotion},
  author={Schaeffer, Paul J and Lindstedt, Stan L},
  journal={Compr Physiol},
  volume={3},
  number={1},
  pages={289--314},
  year={2013}
}

@article{Fajen2013,
  title={Guiding locomotion in complex, dynamic environments},
  author={Fajen, Brett R},
  journal={Frontiers in behavioral neuroscience},
  volume={7},
  pages={85},
  year={2013},
  publisher={Frontiers Media SA}
}

@book{Biewener2018,
    author = {Biewener, Andrew and Patek, Sheila},
    title = "{Animal Locomotion}",
    publisher = {Oxford University Press},
    year = {2018},
    month = {04},
    isbn = {9780198743156},
    doi = {10.1093/oso/9780198743156.001.0001}
}

@article{Volpe_2025,
year = {2025},
month = {aug},
publisher = {IOP Publishing},
volume = {37},
number = {33},
pages = {333501},
author = {Volpe, Giorgio and others}, 
title = {Roadmap for animate matter},
journal = {Journal of Physics: Condensed Matter}
}

@ARTICLE{snake1,
  author={Takemori, Tatsuya and Tanaka, Motoyasu and Matsuno, Fumitoshi},
  journal={IEEE Transactions on Robotics}, 
  title={Gait Design for a Snake Robot by Connecting Curve Segments and Experimental Demonstration}, 
  year={2018},
  volume={34},
  number={5},
  pages={1384-1391},
  doi={10.1109/TRO.2018.2830346}}

@ARTICLE{snake2,
  author={Tanaka, Motoyasu and Tanaka, Kazuo and Matsuno, Fumitoshi},
  journal={IEEE/ASME Transactions on Mechatronics}, 
  title={Approximate Path-Tracking Control of Snake Robot Joints With Switching Constraints}, 
  year={2015},
  volume={20},
  number={4},
  pages={1633-1641},
  doi={10.1109/TMECH.2014.2367657}}

@ARTICLE{snake3,
  author={Fu Qiyuan and Li Chen},
  journal={R. Soc. Open Sci},
title={Robotic modelling of snake traversing large, smooth obstacles reveals stability benefits of body compliance}, 
  year={2020},
  volume={7},
  pages={7191192}}

@article{snake4,
    author = {Fu, Qiyuan and Gart, Sean W and Mitchel, Thomas W and Kim, Jin Seob and Chirikjian, Gregory S and Li, Chen},
    title = {Lateral Oscillation and Body Compliance Help Snakes and Snake Robots Stably Traverse Large, Smooth Obstacles},
    journal = {Integrative and Comparative Biology},
    volume = {60},
    number = {1},
    pages = {171-179},
    year = {2020},
    month = {03},
    issn = {1540-7063}
}

@article{MOOSAVI2025103418,
title = {Snake robots: A state-of-the-art review on design, locomotion, control, and real-world applications},
journal = {Mechatronics},
volume = {112},
pages = {103418},
year = {2025},
issn = {0957-4158},
author = {Syed Kumayl Raza Moosavi and Muhammad Hamza Zafar and Filippo Sanfilippo}
}

@article{Jeger2024,
  title={Adaptive morphing of wing and tail for stable, resilient, and energy-efficient flight of avian-inspired drones},
author={Jeger, Simon Luis and  Wüest, Valentin and  Toumieh, Charbel and  Floreano, Dario},
journal={npj Robot},
 volume={2},
  pages={8},
year={2024},
}

@article{Phan2024,
  title={A twist of the tail in turning maneuvers of bird-inspired drones},
author={Phan, Hoang-Vu and  Floreano, Dario},
journal={Science Robotics },
 volume={9},
  pages={20},
year={2024}
}

@article{Chen2019,
  title={A twist of the tail in turning maneuvers of bird-inspired drones},
author={ Chen, Yufeng and  Zhao, Huichan and  Mao, Jie and  Chirarattananon, Pakpong and  Helbling, E. Farrell and  Hyun, Nak-seung Patrick and Clarke, David R.  and   Wood, Robert J.},
journal={Nature},
 volume={575},
  pages={324},
year={2019}
}

@article{Das2023,
  title={An earthworm-like modular soft robot for locomotion in multi-terrain environments},
author={  Das, Riddhi and  Prashanth, Saravana and   Babu, Murali and  Visentin, Francesco and  Palagi, Stefano and    Mazzolai, Barbara },
journal={Scientific Reports },
 volume={13},
  pages={1571},
year={2023}
}

@article{Gillam2011,
  title={An Introduction to Animal Communication},
author={  Gillam, E. },
journal={Nature Education Knowledge  },
 volume={3},
  pages={20},
year={2011}
}

@article{Gompper2025,
  title={An Introduction to Animal Communication},
author={   Gompper, Gerhard and others}, 
journal={J. Phys.: Cond. Matt   },
 volume={37},
  pages={143501},
year={2025}
}

@book{Hauser1997,
  title={The Evolution of Communication},
  author={Hauser, M.},
  year={1997},
  publisher={Cambridge, MA: MIT Press}
}

@article{Aguilar2016,
  title={A review on locomotion robophysics: the study of movement at the intersection of robotics, soft matter and dynamical systems},
  author={Aguilar, Jeffrey and others}, 
  journal={Reports on Progress in Physics},
  volume={79},
  number={11},
  pages={110001},
  year={2016},
  publisher={IOP Publishing}
}

@Article{Levine2023,
author ="Levine, Herbert and Goldman, Daniel I.",
title  ="Physics of smart active matter: integrating active matter and control to gain insights into living systems",
journal  ="Soft Matter",
year  ="2023",
volume  ="19",
issue  ="23",
pages  ="4204-4207",
publisher  ="The Royal Society of Chemistry",
doi  ="10.1039/D3SM00171G"
}

@article{Wang2021,
  title = {Emergent Field-Driven Robot Swarm States},
  author = {Wang, Gao and others},
  journal = {Phys. Rev. Lett.},
  volume = {126},
  issue = {10},
  pages = {108002},
  numpages = {5},
  year = {2021},
  month = {Mar},
  publisher = {American Physical Society}
}

@article{couzin2005effective,
  title={Effective leadership and decision-making in animal groups on the move},
  author={Couzin, Iain and others},
  journal={Nature},
  volume={433},
  number={7025},
  pages={513--516},
  year={2005},
  publisher={Nature Publishing Group UK London}
}

@Article{Lyu2023,
AUTHOR = {Lyu, Mingyang and Zhao, Yibo and Huang, Chao and Huang, Hailong},
TITLE = {Unmanned Aerial Vehicles for Search and Rescue: A Survey},
JOURNAL = {Remote Sensing},
VOLUME = {15},
YEAR = {2023},
NUMBER = {13},
ARTICLE-NUMBER = {3266}
}

@article{ecke2022uav,
  title={UAV-based forest health monitoring: A systematic review},
  author={Ecke, Simon and Dempewolf, Jan and Frey, Julian and Schwaller, Andreas and Endres, Ewald and Klemmt, Hans-Joachim and Tiede, Dirk and Seifert, Thomas},
  journal={Remote Sensing},
  volume={14},
  number={13},
  pages={3205},
  year={2022},
  publisher={MDPI}
}

@article{husman2021unmanned,
  title={Unmanned aerial vehicles for crowd monitoring and analysis},
  author={Husman, Muhammad Afif and others},
  journal={Electronics},
  volume={10},
  number={23},
  pages={2974},
  year={2021},
  publisher={MDPI}
}

@inproceedings{conte2021performance,
  title={Performance analysis for human crowd monitoring to control covid-19 disease by drone surveillance},
  author={Conte, Claudia and de Alteriis, Giorgio and De Pandi, Francesco and Caputo, Enzo and Moriello, Rosario Schiano Lo and Rufino, Giancarlo and Accardo, Domenico},
  booktitle={2021 IEEE 8th International Workshop on Metrology for AeroSpace (MetroAeroSpace)},
  pages={31--36},
  year={2021},
  organization={IEEE}
}

@article{al2017crowd,
  title={Crowd monitoring system using unmanned aerial vehicle (UAV)},
  author={Al-Sheary, Ali and Almagbile, Ali},
  journal={Journal of Civil Engineering and Architecture},
  volume={11},
  number={11},
  pages={1014--1024},
  year={2017}
}

@article{gohari2022involvement,
  title={Involvement of surveillance drones in smart cities: A systematic review},
  author={Gohari, Adel and Ahmad, Anuar Bin and Rahim, Ruzairi Bin Abdul and Supa’at, Abu Sahmah M and Abd Razak, Shukor and Gismalla, Mohammed Salih Mohammed},
  journal={IEEE Access},
  volume={10},
  pages={56611--56628},
  year={2022},
  publisher={IEEE}
}

@article{Marchetti2013,
  title = {Hydrodynamics of soft active matter},
  author = {Marchetti, M. C. and Joanny, J. F. and Ramaswamy, S. and Liverpool, T. B. and Prost, J. and Rao, Madan and Simha, R. Aditi},
  journal = {Rev. Mod. Phys.},
  volume = {85},
  issue = {3},
  pages = {1143--1189},
  numpages = {0},
  year = {2013},
  month = {Jul},
  publisher = {American Physical Society},
  doi = {10.1103/RevModPhys.85.1143},
  url = {https://link.aps.org/doi/10.1103/RevModPhys.85.1143}
}

@article{marshall2024aims,
  title={On aims and methods of collective animal behaviour},
  author={Marshall, James AR and Reina, Andreagiovanni},
  journal={Animal Behaviour},
  volume={210},
  pages={189--197},
  year={2024},
  publisher={Elsevier}
}

@article{portugal2014upwash,
  title={Upwash exploitation and downwash avoidance by flap phasing in ibis formation flight},
  author={Portugal, Steven J and Hubel, Tatjana Y and Fritz, Johannes and Heese, Stefanie and Trobe, Daniela and Voelkl, Bernhard and Hailes, Stephen and Wilson, Alan M and Usherwood, James R},
  journal={Nature},
  volume={505},
  number={7483},
  pages={399--402},
  year={2014},
  publisher={Nature Publishing Group UK London}
}

@article{reebs2000can,
  title={Can a minority of informed leaders determine the foraging movements of a fish shoal?},
  author={Reebs, Stephan G},
  journal={Animal behaviour},
  volume={59},
  number={2},
  pages={403--409},
  year={2000},
  publisher={Elsevier}
}

@article{swaney2001familiarity,
  title={Familiarity facilitates social learning of foraging behaviour in the guppy},
  author={Swaney, Will and Kendal, Jeremy and Capon, Hannah and Brown, Culum and Laland, Kevin N},
  journal={Animal Behaviour},
  volume={62},
  number={3},
  pages={591--598},
  year={2001},
  publisher={Elsevier}
}

@article{MarchPons2025,
  author  = {March-Pons, David and Pastor-Satorras, Romualdo and Miguel, M. Carmen},
  title   = {Non-linear inhibitory responses enhance performance in collective decision-making},
  journal = {Communications Physics},
  volume  = {8},
  pages   = {119},
  year    = {2025},
  doi     = {10.1038/s42005-025-02046-9},
}

@article{MarchPonsEtAl2024,
  author  = {March-Pons, David and Ferrero, Ezequiel E. and Miguel, M. Carmen},
  title   = {Consensus formation and relative stimulus perception in quality-sensitive, interdependent agent systems},
  journal = {Physical Review Research},
  volume  = {6},
  pages   = {043205},
  year    = {2024},
  doi     = {10.1103/PhysRevResearch.6.043205},
}

@article{AcostaSobaEtAl2025,
  author       = {Acosta-Soba, Daniel and Columbu, Alessandro and Viglialoro, Giuseppe},
  title        = {Boundedness in a nonlinear chemotaxis-consumption model with gradient terms},
  journal      = {arXiv preprint arXiv:2501.13224},
  year         = {2025},
}

@article{AbdoliSharmaLowen2025,
  title={Enhanced efficiency in shear-loaded Brownian gyrators},
  author={Abdoli, Iman and Sharma, Abhinav and L{\"o}wen, Hartmut},
  journal={Physics of Fluids},
  volume={37},
  number={4},
  year={2025},
  publisher={AIP Publishing}
}

@article{lowen2025towards,
  title={Towards intelligent active particles},
  author={L{\"o}wen, Hartmut and Liebchen, Benno},
  journal={arXiv preprint arXiv:2501.08632},
  year={2025}
}

@article{WetzelEtAl2025,
  author        = {Wetzel, Sebastian Johann and Ha, Seungwoong and Iten, Raban and Klopotek, Miriam and Liu, Ziming},
  title         = {Interpretable Machine Learning in Physics: A Review},
  journal       = {arXiv preprint arXiv:2503.23616},
  year          = {2025}
}

@article{JungEtAl2025,
  author  = {Jung, Gerhard and others},
  title   = {Roadmap on machine learning glassy dynamics},
  journal = {Nature Reviews Physics},
  volume  = {7},
  number  = {2},
  pages   = {91--104},
  year    = {2025},
  doi     = {10.1038/s42254-024-00791-4},
}

@article{CichosEtAl2020,
  author  = {Cichos, Frank and Gustavsson, Kristian and Mehlig, Bernhard and Volpe, Giovanni},
  title   = {Machine learning for active matter},
  journal = {Nature Machine Intelligence},
  volume  = {2},
  number  = {2},
  pages   = {94--103},
  year    = {2020},
  doi     = {10.1038/s42256-020-0146-9},
}

@article{cai2025reinforcement,
    author = {Cai, Wenjie and Wang, Gongyi and Zhang, Yu and Qu, Xiang and Huang, Zihan},
    title = {Reinforcement learning for active matter},
    journal = {Biophysics Reviews},
    volume = {6},
    number = {3},
    pages = {031302},
    year = {2025},
    month = {09},
    issn = {2688-4089}
}

@book{sutton1998reinforcement,
  title     = {Reinforcement Learning: An Introduction},
  author    = {Sutton, Richard S. and Barto, Andrew G.},
  year      = {1998},
  publisher = {MIT Press},
  address   = {Cambridge, MA},
}

@article{te2024review,
  title={Metareview: a survey of active matter reviews},
  author={Te Vrugt, Michael and Wittkowski, Raphael},
  journal={The European Physical Journal E},
  volume={48},
  number={3},
  pages={12},
  year={2025},
  publisher={Springer}
}

@article{janzen2023dead,
  title={Dead or alive: Distinguishing active from passive particles using supervised learning (a)},
  author={Janzen, Giulia and others}, 
  journal={Europhysics Letters},
  volume={143},
  number={1},
  pages={17004},
  year={2023},
  publisher={IOP Publishing}
}

@article{braat2025shape,
  title={Shape matters: Inferring the motility of confluent cells from static images},
  author={Braat, Quirine  and Janzen, Giulia and Jansen, Bas C and Debets, Vincent Emanuel and Ciarella, Simone and Janssen, Liesbeth},
  journal={Soft Matter},
  year={2025},
  publisher={Royal Society of Chemistry}
}

@article{janzen2024classifying,
  title={Classifying the age of a glass based on structural properties: A machine learning approach},
  author={Janzen, Giulia and others},
  journal={Physical Review Materials},
  volume={8},
  number={2},
  pages={025602},
  year={2024},
  publisher={APS}
}

@book{brunton2022data, place={Cambridge}, edition={2, Cambridge University Press}, title={Data-Driven Science and Engineering: Machine Learning, Dynamical Systems, and Control}, publisher={Cambridge University Press}, author={Brunton, Steven L. and Kutz, J. Nathan}, year={2022}}

@article{bechinger2016active,
  title={Active particles in complex and crowded environments},
  author={Bechinger, Clemens and Di Leonardo, Roberto and L{\"o}wen, Hartmut and Reichhardt, Charles and Volpe, Giorgio and Volpe, Giovanni},
  journal={Reviews of modern physics},
  volume={88},
  number={4},
  pages={045006},
  year={2016},
  publisher={APS}
}

@article{wang2015one,
  title={From one to many: Dynamic assembly and collective behavior of self-propelled colloidal motors},
  author={Wang, Wei and Duan, Wentao and Ahmed, Suzanne and Sen, Ayusman and Mallouk, Thomas E},
  journal={Accounts of chemical research},
  volume={48},
  number={7},
  pages={1938--1946},
  year={2015},
  publisher={ACS Publications}
}

@article{ruiz2024discovering,
  title={Discovering dynamic laws from observations: the case of self-propelled, interacting colloids},
  author={Ruiz-Garcia, Miguel and Barriuso G, C Miguel and Alexander, Lachlan C and Aarts, Dirk GAL and Ghiringhelli, Luca M and Valeriani, Chantal},
  journal={Physical Review E},
  volume={109},
  number={6},
  pages={064611},
  year={2024},
  publisher={APS}
}

@article{fletcher2010cell,
  title={Cell mechanics and the cytoskeleton},
  author={Fletcher, Daniel A and Mullins, R Dyche},
  journal={Nature},
  volume={463},
  number={7280},
  pages={485--492},
  year={2010},
  publisher={Nature Publishing Group UK London}
}

@article{grada2017research,
  title={Research techniques made simple: analysis of collective cell migration using the wound healing assay},
  author={Grada, Ayman and others},
  journal={Journal of Investigative Dermatology},
  volume={137},
  number={2},
  pages={e11--e16},
  year={2017},
  publisher={Elsevier}
}

@article{angelini2011glass,
  title={Glass-like dynamics of collective cell migration},
  author={Angelini, Thomas and others},
  journal={Proceedings of the National Academy of Sciences},
  volume={108},
  number={12},
  pages={4714--4719},
  year={2011},
  publisher={National Acad Sciences}
}

@article{bi2016motility,
  title={Motility-driven glass and jamming transitions in biological tissues},
  author={Bi, Dapeng and Yang, Xingbo and Marchetti, M Cristina and Manning, M Lisa},
  journal={Physical Review X},
  volume={6},
  number={2},
  pages={021011},
  year={2016},
  publisher={APS}
}

@article{clegg2021characterising,
  title={Characterising soft matter using machine learning},
  author={Clegg, Paul S},
  journal={Soft Matter},
  volume={17},
  number={15},
  pages={3991--4005},
  year={2021},
  publisher={Royal Society of Chemistry}
}

@article{bapst2020unveiling,
  title={Unveiling the predictive power of static structure in glassy systems},
  author={Bapst, Victor and others}, 
  journal={Nature physics},
  volume={16},
  number={4},
  pages={448--454},
  year={2020},
  publisher={Nature Publishing Group UK London}
}

@article{boattini2021averaging,
  title={Averaging local structure to predict the dynamic propensity in supercooled liquids},
  author={Boattini, Emanuele and Smallenburg, Frank and Filion, Laura},
  journal={Physical Review Letters},
  volume={127},
  number={8},
  pages={088007},
  year={2021},
  publisher={APS}
}

@article{minartz2025deep,
  title={Deep Neural Cellular Potts Models},
  author={Minartz, Koen and others},
  journal={arXiv preprint arXiv:2502.02129},
  year={2025}
}

@article{liu2025machine,
  title={Machine Learning-Based Methods for Materials Inverse Design: A Review.},
  author={Liu, Yingli and others},
  journal={Computers,Materials\&Continua},
  volume={82},
  number={2},
  year={2025}
}

@article{PhysRevResearch.3.033291,
  title = {Learning to control active matter},
  author = {Falk, Martin J. and others},
  journal = {Phys. Rev. Res.},
  volume = {3},
  issue = {3},
  pages = {033291},
  numpages = {12},
  year = {2021},
  month = {Sep},
  publisher = {American Physical Society}
}

@article{lieu2025dynamic,
  title={Dynamic control of self-assembly of quasicrystalline structures through reinforcement learning},
  author={Lieu, Uyen Tu and Yoshinaga, Natsuhiko},
  journal={Soft Matter},
  volume={21},
  number={3},
  pages={514--525},
  year={2025},
  publisher={Royal Society of Chemistry}
}

@article{floyd2024learning,
  title={Learning to control non-equilibrium dynamics using local imperfect gradients},
  author={Floyd, Carlos and others},
  journal={arXiv preprint arXiv:2404.03798},
  year={2024}
}

@article{baulin2025intelligent,
  title={Intelligent soft matter: towards embodied intelligence},
  author={Baulin, Vladimir A and others},
  journal={Soft Matter},
  volume={21},
  number={21},
  pages={4129--4145},
  year={2025},
  publisher={Royal Society of Chemistry}
}

@article{JuACSnano2025,
  title={Technology Roadmap of Micro/Nanorobots},
        author={ Ju,  Xiaohui and others },    
            journal={ACS Nano},
            volume={19}, 
             pages={24174},
  year={2025}}

@InCollection{volpebook2024,
  author    = {Volpe, Giorgio and Cichos, Frank and Volpe, Giovanni},
  title     = {Active Matter and Artificial Intelligence},
  booktitle = {Active Colloids: From Fundamentals to Frontiers},
  publisher = {Royal Society of Chemistry},
  year      = {2024},
  pages     = {565--577},
}

@article{volpeworm2025,
  title={Inchworm-inspired soft robot locomotion based on groove-guided locomotion},
  author={ Thanabalan, Hari Prakash and  Bengtsson, Lars and   Lafont, Ugo and  Volpe, Giovanni},
  journal={Proc. SPIE PC13585, Emerging Topics in Artificial Intelligence (ETAI)},
  year={2025}
}

@INPROCEEDINGS{GenPath,
  author={Bartolini, Novella and Coletta, Andrea and Maselli, Gaia and Piva, Mauro and Silvestri, Domenicomichele},
  booktitle={IEEE INFOCOM 2021 - IEEE Conference on Computer Communications Workshops (INFOCOM WKSHPS)}, 
  title={GenPath - A Genetic Multi-Round Path Planning Algorithm for Aerial Vehicles}, 
  year={2021},
  volume={},
  number={},
  pages={1-6}
}

@Misc{prorok2021,
  author        = {Amanda Prorok and Matthew Malencia and Luca Carlone and Gaurav S. Sukhatme and Brian M. Sadler and Vijay Kumar},
  title         = {Beyond Robustness: A Taxonomy of Approaches towards Resilient Multi-Robot Systems},
  year          = {2021},
  archiveprefix = {arXiv},
  eprint        = {2109.12343},
  primaryclass  = {cs.RO},
  url           = {https://arxiv.org/abs/2109.12343},
}

@article{yao2022guiding,
  author ={Yao, Weijia and de Marina, H{\'e}ctor Garc{\'\i}a and Sun, Zhiyong and Cao, Ming},
  title ={Guiding vector fields for the distributed motion coordination of mobile robots},
  journal={IEEE Transactions on Robotics},
  volume={39},
  number={2},
  pages={1119--1135},
  year={2022},
  publisher={IEEE}
}

@Article{Gielis2022,
  author  = {Gielis, J. and Shankar, A. and Prorok, A.},
  title   = {A Critical Review of Communications in Multi-robot Systems},
  journal = {Current Robotics Reports},
  year    = {2022},
  volume  = {3},
  pages   = {213--225}
}

@Article{Kegeleirs2025,
  author    = {Miquel Kegeleirs and Mauro Birattari},
  journal   = {Frontiers Robotics {AI}},
  title     = {Towards applied swarm robotics: current limitations and enablers},
  year      = {2025},
  volume    = {12},
  bibsource = {dblp computer science bibliography, https://dblp.org},
  doi       = {10.3389/FROBT.2025.1607978},
}

@InProceedings{mondada2009puck,
  author       = {Mondada, Francesco and others},
  title        = {The e-puck, a robot designed for education in engineering},
  booktitle    = {Proceedings of the 9th conference on autonomous robot systems and competitions},
  year         = {2009},
  volume       = {1},
  pages        = {59--65},
  organization = {Instituto Polit{\'e}cnico de Castelo Branco (IPCB), Castelo Branco},
}

@inproceedings{rubenstein2012kilobot,
  title={Kilobot: A low cost scalable robot system for collective behaviors},
  author={Rubenstein, Michael and Ahler, Christian and Nagpal, Radhika},
  booktitle={2012 IEEE international conference on robotics and automation},
  pages={3293--3298},
  year={2012},
  organization={IEEE}
}

@inproceedings{giernacki2017crazyflie,
  title={Crazyflie 2.0 quadrotor as a platform for research and education in robotics and control engineering},
  author={Giernacki, Wojciech and others},
  booktitle={2017 22nd international conference on methods and models in automation and robotics (MMAR)},
  pages={37--42},
  year={2017},
  organization={IEEE}
}

@INPROCEEDINGS{paull,
  author={Paull, Liam and others},
  booktitle={2017 IEEE International Conference on Robotics and Automation (ICRA)}, 
  title={Duckietown: An open, inexpensive and flexible platform for autonomy education and research}, 
  year={2017},
  volume={},
  number={},
  pages={1497-1504},
  doi={10.1109/ICRA.2017.7989179}
}

@ARTICLE{wilson,
  author={Wilson, Sean and Glotfelter, Paul and Wang, Li and Mayya, Siddharth and Notomista, Gennaro and Mote, Mark and Egerstedt, Magnus},
  journal={IEEE Control Systems Magazine}, 
  title={The Robotarium: Globally Impactful Opportunities, Challenges, and Lessons Learned in Remote-Access, Distributed Control of Multirobot Systems}, 
  year={2020},
  volume={40},
  number={1},
  pages={26-44},
  doi={10.1109/MCS.2019.2949973}
}

@InProceedings{garcia-perez,
  author    = {Garc{\'i}a-P{\'e}rez, L{\'i}a
and Sombr{\'i}a, Jes{\'u}s Chac{\'o}n
and Font{\'a}n, Alejandro Gutierrez
and Castellanos, Juan Francisco Jim{\'e}nez},
  booktitle = {Robotics in Education},
  title     = {Collaborative Construction of a Multi-Robot Remote Laboratory: Description and Experience},
  year      = {2023},
  address   = {Cham},
  editor    = {Balogh, Richard
and Obdr{\v{z}}{\'a}lek, David
and Christoforou, Eftychios},
  pages     = {243--254},
  publisher = {Springer Nature Switzerland},
  isbn      = {978-3-031-38454-7},
}

@InCollection{sakurama,
  author    = {Sakurama, Kazunori and Sugie, Toshiharu},
  title     = {<chapter/paper title>},
  booktitle = {Generalized Coordination of Multi-robot Systems},
  year      = {2021},
  pages     = {<start>--<end>},
  publisher = {<publisher>},
  editor    = {<editor(s)>},
}

@ARTICLE{10092576,
  author={Ma\~{n}as-Álvarez, Francisco-José and others},
  journal={IEEE Access}, 
  title={Robotic Park: Multi-Agent Platform for Teaching Control and Robotics}, 
  year={2023},
  volume={11},
  number={},
  pages={34899-34911},
  doi={10.1109/ACCESS.2023.3264508}
}

@Article{mokhtarian,
  author   = {Mokhtarian, Armin and others},
  title    = {A Survey on Small-Scale Testbeds for Connected and Automated Vehicles and Robot Swarms: A Guide for Creating a New Testbed [Survey]},
  journal  = {IEEE Robotics \& Automation Magazine},
  year     = {2025},
  volume   = {32},
  number   = {3},
  pages    = {146-163},
  doi      = {10.1109/MRA.2024.3505772},
}

@article{Falko,
author = {Dressler, Falko and Akan, Ozgur B.},
title = {A survey on bio-inspired networking},
year = {2010},
issue_date = {April, 2010},
publisher = {Elsevier North-Holland, Inc.},
address = {USA},
volume = {54},
number = {6},
issn = {1389-1286},
journal = {Comput. Netw.},
month = apr,
pages = {881–900}
}

@Article{brooks1998fast,
  author  = {Brooks, R and Flynn, A},
  journal = {Journal of the British Interplanetary Society},
  title   = {Fast, Cheap and out of Control: A Robot Invasion of the Solar System. S. 478-485},
  year    = {1998},
  volume  = {42},
}

@Article{bredeche2018embodied,
  author    = {Bredeche, Nicolas and Haasdijk, Evert and Prieto, Abraham},
  journal   = {Frontiers in Robotics and AI},
  title     = {Embodied evolution in collective robotics: a review},
  year      = {2018},
  pages     = {12},
  volume    = {5},
  publisher = {Frontiers Media SA},
}

@inproceedings{Mirhosseini2022,
author = {Mirhosseini, Yoones and others},
title = {Adaptive phototaxis of a swarm of mobile robots using positive and negative feedback self-alignment},
year = {2022},
isbn = {9781450392372},
publisher = {Association for Computing Machinery},
address = {New York, NY, USA},
doi = {10.1145/3512290.3528816},
booktitle = {Proceedings of the Genetic and Evolutionary Computation Conference},
pages = {104–112},
numpages = {9},
location = {Boston, Massachusetts},
series = {GECCO '22}
}

@article{BenZion2023,
author = {Matan Yah Ben Zion and others},
title = {Morphological computation and decentralized learning in a swarm of sterically interacting robots},
journal = {Science Robotics},
volume = {8},
number = {75},
pages = {eabo6140},
year = {2023},
doi = {10.1126/scirobotics.abo6140},
eprint = {https://www.science.org/doi/pdf/10.1126/scirobotics.abo6140}}

@article{Deblais2018PRL,
  title = {Boundaries Control Collective Dynamics of Inertial Self-Propelled Robots},
  author = {Deblais, A. and others},
  journal = {Phys. Rev. Lett.},
  volume = {120},
  issue = {18},
  pages = {188002},
  numpages = {5},
  year = {2018},
  month = {May},
  publisher = {American Physical Society},
  doi = {10.1103/PhysRevLett.120.188002},
}

@article{Boudetscience2021,
author = {Boudet, J. F. and others},
title = {From collections of independent, mindless robots to flexible, mobile, and directional superstructures},
journal = {Science Robotics},
volume = {6},
number = {56},
pages = {eabd0272},
year = {2021},
doi = {10.1126/scirobotics.abd0272},
eprint = {https://www.science.org/doi/pdf/10.1126/scirobotics.abd0272}}

@article{Arbel2025,
  author  = {Arbel, Eden and others},
  title   = {A mechanical route for cooperative transport in autonomous robotic swarms},
  journal = {Nature Communications},
  year    = {2025},
  volume  = {16},
  number  = {1},
  pages   = {7519},
  doi     = {10.1038/s41467-025-61896-7},
  issn    = {2041-1723},
  date    = {2025-09-02}
}

@article{CasiulisPNAS2025,
author = {Mathias Casiulis  and others},
title = {A geometric condition for robot-swarm cohesion and cluster–flock transition},
journal = {Proceedings of the National Academy of Sciences},
volume = {122},
number = {37},
pages = {e2502211122},
year = {2025},
doi = {10.1073/pnas.2502211122},
eprint = {https://www.pnas.org/doi/pdf/10.1073/pnas.2502211122}}

@article{alvarez2025segregation,
  title={Segregation and cooperation in active colloidal binary mixtures},
  author={Alvarez, Laura and others},
  journal={arXiv preprint arXiv:2506.15188},
  year={2025}
}

@article{BenZion2022NAtureComm,
  author  = {Ben Zion, Matan Yah and Caba, Yaelin and Modin, Alvin and Chaikin, Paul M.},
  title   = {Cooperation in a fluid swarm of fuel-free micro-swimmers},
  journal = {Nature Communications},
  year    = {2022},
  volume  = {13},
  number  = {1},
  pages   = {184},
  doi     = {10.1038/s41467-021-27870-9},
  issn    = {2041-1723},
  date    = {2022-01-10}
}

@article{DasPRX2024,
  title = {Flocking by Turning Away},
  author = {Das, Suchismita and others},
  journal = {Phys. Rev. X},
  volume = {14},
  issue = {3},
  pages = {031008},
  numpages = {16},
  year = {2024},
  month = {Jul},
  publisher = {American Physical Society},
  doi = {10.1103/PhysRevX.14.031008}
}

@article{ButlerPNAS2010,
author = {Mitchell T. Butler  and Qingfeng Wang  and Rasika M. Harshey },
title = {Cell density and mobility protect swarming bacteria against antibiotics},
journal = {Proceedings of the National Academy of Sciences},
volume = {107},
number = {8},
pages = {3776-3781},
year = {2010},
doi = {10.1073/pnas.0910934107},
eprint = {https://www.pnas.org/doi/pdf/10.1073/pnas.0910934107}}

@article{DunkelPRL2013,
  title = {Fluid Dynamics of Bacterial Turbulence},
  author = {Dunkel, J\"orn and Heidenreich, Sebastian and Drescher, Knut and Wensink, Henricus H. and B\"ar, Markus and Goldstein, Raymond E.},
  journal = {Phys. Rev. Lett.},
  volume = {110},
  issue = {22},
  pages = {228102},
  numpages = {5},
  year = {2013},
  month = {May},
  publisher = {American Physical Society},
  doi = {10.1103/PhysRevLett.110.228102}
}

@article{WensinkPNAS2012,
author = {Henricus H. Wensink  and Jörn Dunkel  and Sebastian Heidenreich  and Knut Drescher  and Raymond E. Goldstein  and Hartmut Löwen  and Julia M. Yeomans },
title = {Meso-scale turbulence in living fluids},
journal = {Proceedings of the National Academy of Sciences},
volume = {109},
number = {36},
pages = {14308-14313},
year = {2012},
eprint = {https://www.pnas.org/doi/pdf/10.1073/pnas.1202032109}}

@article{SokolovPRL2012,
  title   = {Physical Properties of Collective Motion in Suspensions of Bacteria},
  author  = {Sokolov, Andrey and Aranson, Igor S.},
  journal = {Phys. Rev. Lett.},
  volume  = {109},
  number  = {24},
  pages   = {248109},
  year    = {2012},
  month   = dec,
}

\end{document}